\documentclass[preprint,12pt]{aastex}
\newcommand{\degree}{\mbox{$^{\circ}$}}

\newcommand{\as}{\mbox{\arcsec}}

\newcommand{\lsun}{\mbox{L$_\odot$}}% Lsun
\newcommand{\msun}{\mbox{M$_\odot$}}% Msun
\newcommand{\rsun}{\mbox{R$_\odot$}}% Rsun

\def\plotfiddle#1#2#3#4#5#6#7{\centering \leavevmode
\vbox to#2{\rule{0pt}{#2}}
\includegraphics{#1}}

\begin{document}

\slugcomment{v3.0; 14 Oct 2011}

%%%%%%%%%%%%%%%%%% title %%%%%%%%%%%%%%%%%%%%%%%%%%%%%%%%%%%%%%%%
\title {\bf A {\it Herschel} Search For Cold Dust in Brown Dwarf Disks: \quad \quad \quad
First Results}
\author{Paul M. Harvey\altaffilmark{1},
Thomas Henning\altaffilmark{2},
Fran\c cois M\'enard\altaffilmark{3},
Sebastian Wolf\altaffilmark{4},
Yao Liu\altaffilmark{4},
Lucas A. Cieza\altaffilmark{5},
Neal J. Evans II\altaffilmark{1},
Ilaria Pascucci\altaffilmark{6},
Bruno Mer\'in\altaffilmark{7},
Christophe Pinte\altaffilmark{3}
}

\altaffiltext{1}{Astronomy Department, University of Texas at Austin, 1 University Station C1400, Austin, TX 78712-0259;  pmh@astro.as.utexas.edu, nje@astro.as.utexas.edu}
\altaffiltext{2}{Max-Planck-Institut for Astronomy, K\"önigstuhl 17, 69117 Heidelberg, Germany;henning@mpia.de}
\altaffiltext{3}{UJF-Grenoble 1/CNRS-INSU, Institut de Plan\'étologie et d’Astrophysique (IPAG) UMR 5274, BP 53, 38041 Grenoble cedex 9, France; menard@obs.ujf-grenoble.fr, christophe.pinte@obs.ujf-grenoble.fr}
\altaffiltext{4}{University of Kiel, Institute of Theoretical Physics and Astrophysics, Leibnizstr. 15, 24098 Kiel, Germany; wolf@astrophysik.uni-kiel.de, yliu@pmo.ac.cn}
\altaffiltext{5}{Institute for Astronomy, University of Hawaii, 2680 Woodlawn Dr., Honolulu, HI 96822; lcieza@ifa.hawaii.edu, Sagan Fellow}
\altaffiltext{6}{Lunar and Planetary Laboratory, Department of Planetary Sciences, University of Arizona, 1629 E. University Blvd., Tucson, AZ 85721; pascucci@lpl.arizona.edu}
\altaffiltext{7}{Herschel Science Centre, SRE-SDH, ESA P.O. Box 78, 28691 Villanueva de la Ca\~nada (Madrid), Spain; Bruno.Merin@sciops.esa.int}

%%%%%%%%%%%%%%%%%%%% Abstract %%%%%%%%%%%%%%%%%%%%%%%%%%%%%
\begin{abstract}

We report initial results from a {\it Herschel} program to search for far-infrared emission from cold dust around a statistically
significant sample of young brown dwarfs.  The first three objects in our survey are all detected at 70\micron,
and we report the first detection of a brown dwarf at 160\micron.  The flux densities are consistent with the presence of
substantial amounts of cold dust in the outer disks around these objects.  We modeled the SED's
with two different radiative transfer codes.  
We find that a broad range of model parameters provides a reasonable fit to the SED's, but that the addition
of our 70\micron, and especially the 160\micron\ detection enables strong lower limits to be placed on the disk masses
since most of the mass is in the outer disk.  We find likely disk masses in the range of a few $\times 10^{-6}$ to
$10^{-4}$ \msun. Our models provide a good fit to the SED's and do not require dust settling.
%and that it is difficult to constrain any single parameter without making assumptions about others.
%The least massive disk, $\sim$few$\times 10^{-6}$ \msun, resides around ISO138 in
%Cham I, while the two objects in TWA, SSSPM1102 and 2M1207, have substantially more massive disks, $\sim$few$\times 10^{-5}$ \msun.
%Without additional observational constraints, many of the other model parameters are relatively unconstrained, although 
%The strong far-ir emission is consistent with substantial grain growth in these disks.

\end{abstract}

%\keywords{infrared: general --- clouds: star forming regions}
%%%%%%%%%%%%%%%%%%% Main text %%%%%%%%%%%%%%%%%%%%%%%%%%%%

\section{Introduction}\label{intro}

Young brown dwarfs exhibit circum-``stellar'' disk phenomena much like their more massive
counterparts, e.g. \citet{klein03}, \citet{apai04} and \citet{luhman10}.  Although there are quantitative differences 
in physical parameters
for disks around sub-stellar objects such as the gas chemistry and degree of dust processing \citep{pascucci09}, the distribution of properties 
is relatively continuous across the sub-stellar boundary, e.g. \citet{scholz09}.  
Brown dwarfs (BD's) provide a qualitatively and quantitatively different physical environment in which to study
disk structure and evolution, particularly in conditions more favorable to lower mass planet formation and with lower
stellar heating and ionizing fluxes.
Observational selection effects, however, make the study of BD disks more
difficult than for T Tauri stars because of the low luminosities.  The quantity
of cold, T $<$ 150K, dust in the disks is difficult to determine without sensitive measurements at $\lambda >$30\micron.  
There are two reported {\it Spitzer} detections of BD's at 70\micron\ \citep{guieu07,riaz08}, and several
have been detected at $\lambda \sim$ 1 mm \citep{klein03, scholz06} with modest S/N.  But the detected number is painfully
small for any statistical investigation of the cold dust mass, disk flaring, dust emissivity,
grain growth, and other properties.

We describe our earliest results from a {\it Herschel} \citep{pilbrat10} GT1 (Guaranteed Time, Phase 1) program with the PACS photometer \citep{pacs10}
that will eventually provide sensitive
photometry at 70 and 160\micron\ of 50 young BD's in several star-forming regions spanning a range of ages and spectral types.
Our earliest sample is a strong function of {\it Herschel} scheduling and comprises three objects, two in the 
relatively older (8--10 Myr) TW Hya
association (TWA),
and one in the younger ($\sim$ 2 Myr) Chamaeleon I region.  
%In a preliminary attempt to estimate the likely range of possible brightnesses of these
%three, all were expected to be close to our detection limit, roughly 3-4 mJy at 70\micron\ as described in \S\ref{obs}.
%In fact, we detected all three at 70\micron\ with reasonable S/N and one at 160\micron. 
%The observational parameters are described in \S\ref{obs} and the observational results in \S\ref{obsresults}.
% we discuss the observational results
%relative to previous infrared observations.  
%In \S\ref{model} we
%describe our modeling based on the observed spectral energy distributions (SED's) from 1--160\micron,
%and in \S\ref{anal} we discuss these results in comparison to previous models for BD and T Tauri disks.
%Finally in \S\ref{summ} we summarize our conclusions and expected future progress.

\section{Observations and Data Reduction}\label{obs}

%Our program, GT1\_pharve01\_2, takes advantage of the excellent sensitivity and spatial
%sampling of the PACS photometer on {\it Herschel}.  
%All the observations use the same mode
%with essentially the same total integration time.   
All the observations used the PACS ``mini-scan-map'' mode with similar
integration times. This mode simultaneously images at 
70 and 160\micron\ with two scan maps at the
recommended relative angle for good 1/f noise reduction and high sensitivity over an area 
$\sim$ 60\arcsec$\times$90\arcsec.  Table \ref{obstable} lists the relevant parameters 
and Table \ref{aortable} lists the AOR's.  
%The 1$\sigma$ noise for the combined AOR's 
%was estimated at $\sim$ 1 mJy at 70\micron\ and 2.3 mJy at 160\micron\ by the
%{\it Herschel} observation-planning tool HSPOT.  In fact, we achieved slightly better noise levels,
%probably because of our use of psf-fitting photometry instead of aperture photometry as discussed below.

\begin{figure}
\plotfiddle{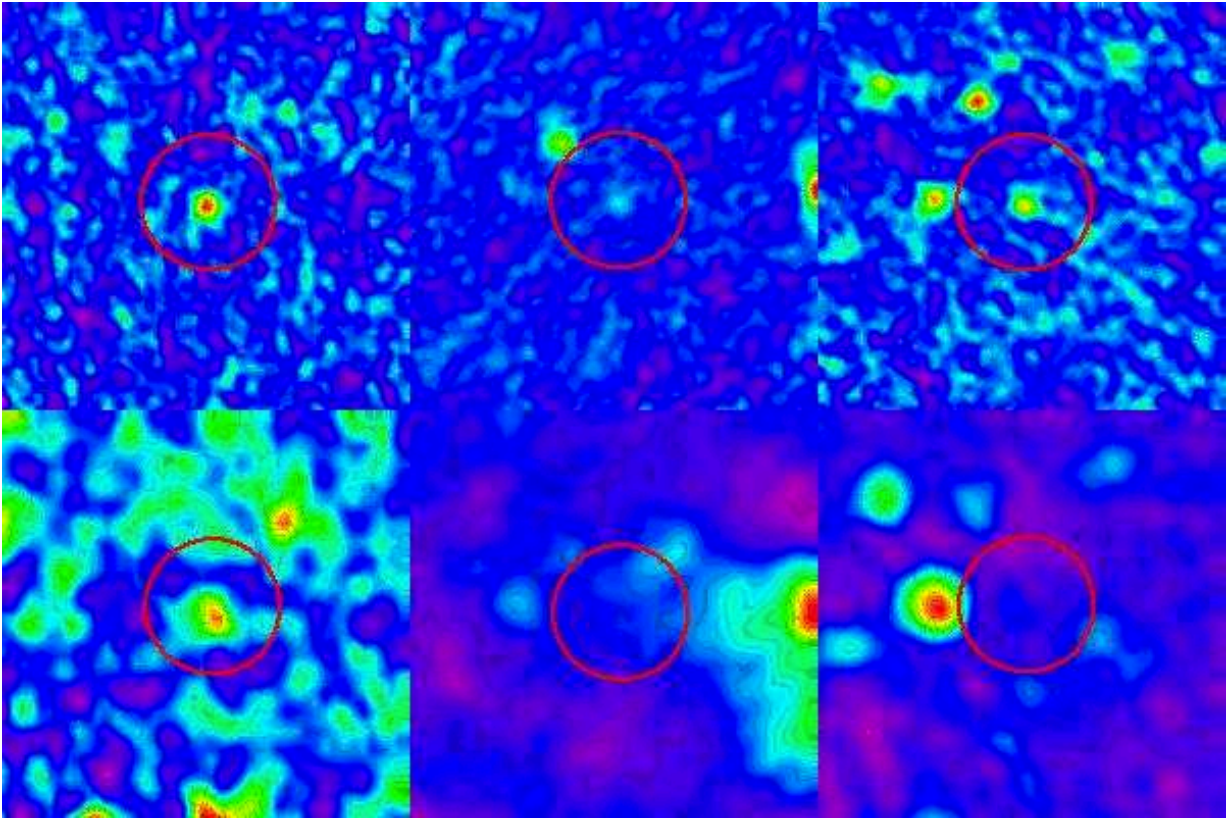}{5.0in}{-90}{60}{60}{-250}{400}
\figcaption{\label{obsfigs}
Top, left to right - 70\micron\ images, 90\arcsec\ square: SSSPM1102, ISO138, and 2MASS1207;
bottom - 160\micron\ images.  Circles are 30\arcsec\ diameter centered on the nominal source
positions.}
\end{figure}

The data were first processed with the Herschel Interactive Processing Environment version 7.0, with 
standard high-pass filtering for point-source observations to produce %, and which processes the
%two separate AOR's for each object together.  
%The portions of scan legs where the telescope was stopped or not
%scanning at a uniform velocity were not used for this processing.  The output of this script consists
fits files of the image, coverage, and
uncertainty.  The uncertainty images are not yet reliable, so we estimated uncertainties
as described below.  Figure \ref{obsfigs} shows the final mosaicked images.  We also
processed each AOR separately to test that the individual observations were consistent with the
combined set for each object, i.e. the
expected $\sqrt 2$ reduction in S/N.

The final processing utilized the psf-fitting photometry tool, {\it c2dphot}, from the
{\it c2d} {\it Spitzer} Legacy Team \citep{harvey06,evans07}, and based on the earlier DOPHOT tool \citep{dophot}.   
This tool can be used in various modes,
including finding peaks above the background and fitting a psf to the local maxima, or fitting a psf to a fixed
position, a mode useful for estimating noise and determining upper limits.  Our quoted noise 
was determined this way by fitting a half-dozen arbitrarily chosen, nominally
empty parts of the image within the high-coverage area.  Upper limits for the two objects undetected
at 160\micron\ were also determined this way, and also by inserting artificial sources in the images at various
flux levels. This tool produces aperture fluxes for comparison with the psf-fitted values, and
those values agree well.  In addition, aperture-flux ``curves-of-growth'' show
good agreement with the psf-fit photometry.

\section{Observational Results}\label{obsresults}

Figure \ref{obsfigs} shows all three objects clearly detected at 70\micron\ and 
SSSPM1102 also at 160\micron.   The level of structured background at 160\micron\ is higher
for the other two BD's; the upper limits at 160\micron\ for those two were set
by this structure rather than by the instrumental sensitivity and DC sky background.  There {\it is} some
weak, diffuse emission at 160\micron\ close to the position of 2M1207, but the offset from both
the nominal source position and the 70\micron\ centroid is probably too large to be consistent with emission
from the object.   In addition to our confidence in these detections based on inspection
of the images, we note that {\it c2dphot} reliably extracted the sources
as bona fide point sources with no confusion in its most general ``source-finding'' mode. 
The small offset from the nominal position for SSSPM1102, $\sim$ 2\as, is essentially identical
in both wavelength channels and well within the typical {\it Herschel} pointing uncertainty.

\begin{figure}
%\plotfiddle{f2.eps}{6.5in}{0}{70}{70}{-250}{-15}
\plotfiddle{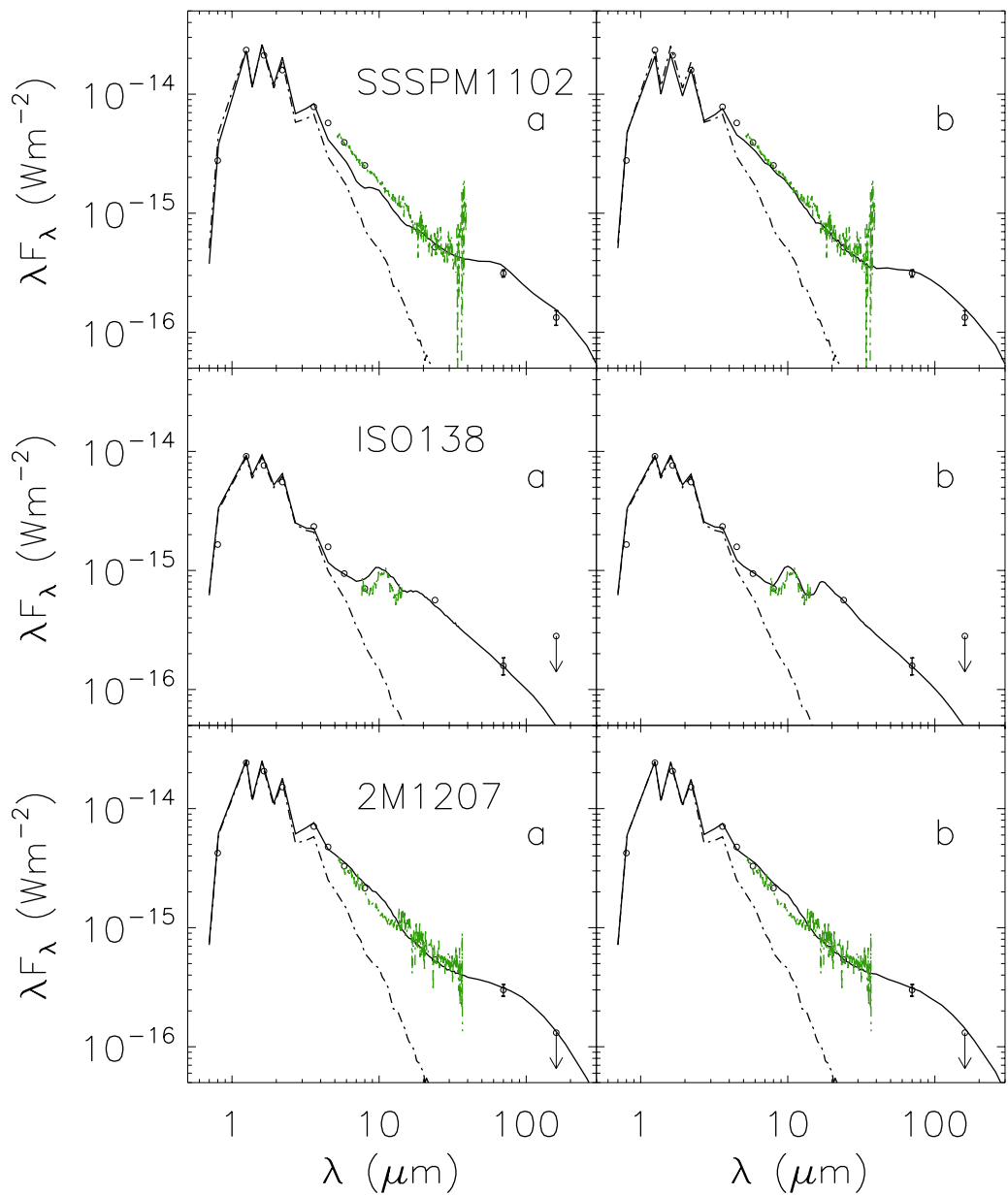}{6.5in}{0}{70}{70}{-250}{-15}
\figcaption{\label{sedfig}
SED's of the three observed BD's (open circles).  {\it Spitzer} IRS data are
shown in green.  Solid lines show SED's from the example fitted models listed in Table \ref{modparms}.
Dashed lines show the bare sub-stellar photospheres used in
the model, extincted by the chosen $A_v$.}
\end{figure}

The derived flux densities and
upper limits are listed in Table \ref{aortable}.  All three objects have also been observed
by {\it Spitzer} with IRAC and MIPS \citep{riaz06,riaz08,luhman10}.  Two
were observed with the IRS instrument over its whole spectral range \citep{riaz07, riaz08, morrow08}, and one (ISO138)
over the 10\micron\ silicate feature \citep{pascucci09}.   All are detected in the 2MASS catalog
and have reasonably well-determined spectral types as described below.  We have
collected these and available shorter wavelength data into the SED's in
Figure \ref{sedfig}.

%Discuss SED shape, obvious results....
Figure \ref{sedfig} shows that the SED's of the two BD's in the TW Hya association are much more similar to each other than
to ISO138 in Cham I.  In particular, the [8]--[24] color and to a lesser extent the [8]--[70] color of ISO138 is
substantially redder than for the other two.  The two TWA BD's also show strong excess emission to
wavelengths as short as the IRAC 3.6\micron\ band.   Most importantly, all clearly show emission in 
the far-infrared indicative of substantial amounts of cold dust.
%to the longest wavelengths where flux is detected.  Next we describe our initial 
%modeling of these circum-object disks and the constraints that the models place on the disk properties.

\section{Modeling}\label{model}

%Discuss briefly the characteristics of the two different codes, put results in a Table?, discuss
%level of confidence and error bars on derived quantities.

We have modeled the SED's of these three BD's with two different codes, MC3D described by \citet{wolf99, wolf03}
and MCFOST described by \citet{pinte06, pinte09}.  Both codes are three-dimensional, radiative transfer
codes using the Monte-Carlo method and NextGen stellar atmosphere parameters.
%These modeling results should be viewed as very preliminary and only indicative
%of the direction we plan to take with more complete modeling when we have a statistically significant sample of
%objects observed.  
The only significant differences in the modeling with each code were the range of parameters examined and
a few of the initial assumptions described below.
The goal of this modeling was to understand what disk parameters we are most likely to
be able to constrain in the future, and what assumptions are most critical to define, perhaps by future observations.
%We have, in particular, only tried to fit approximately the part of the SED defined by the silicate features at 10--25\micron,
%since our new data do not bear directly on the detailed silicate mineralogy, and the shape of this feature does
%not significantly impact the derived thermal balance in the outer disk.
For example, we did not attempt to fit the exact shape of the silicate feature, hence the 
exact silicate mineralogy of the dust in the disk atmosphere.   The integrated
flux in the model feature is close to the data,  and the thermal budget of the disk is therefore also correct and sufficiently
accurate for this study.
More detailed modeling will follow when our full BD sample is available.
%When we have the full statistical sample of observations, we will conduct more detailed 
%modeling of all the BD's in the sample.

Both modeling codes parametrize the disk structure geometrically in similar ways.  
Typical parameters that were fixed in our initial modeling include: the
slope of the grain size distribution, $ dn(a) \propto a^{-3.5} da $, minimum grain size, and 
the dust grain properties, typically astronomical silicates as 
described by \citet{drainelee84} with varying amounts of amorphous carbon.
In the MC3D models the surface density power law was
set to $\Sigma(r) \propto r^{-p}$ with $p = 1$, while with MCFOST a range for the exponent from 0.1 to 1.5 was explored, 
depending on the object.
The stellar parameters were fixed based on previous studies referenced below, but models with some variation in parameters 
were tested because of 
uncertainties in spectral types, luminosities, and stellar radii. %that exist due to observational and stellar model uncertainties.  
Typical parameters that differed between models included: maximum grain size, total gas$+$dust disk mass (with an
assumed gas-to-dust mass ratio of 100), inclination angle, inner radius, and the disk scale height
parameters, $h_0$ and $\gamma$, $h(r) = h_0(r/r_0)^\gamma$.  The outer radius was typically fixed at 50--200 AU, but the SED's are
quite insensitive to this choice, as illustrated below.
%; for some models also the surface density power law was allowed to vary with exponents between -1.5 and -0.5.  

%We examined Bayesian probabilities for the combination of the parameters that
%were allowed to vary to determine our confidence level in the fitted parameters.  In general, many of the probability distributions
%were rather flat, and/or showed a substantial correlation with other parameter probability distributions, implying either
%unconstrained fits or fits where only a combination of parameters could be constrained.

For each object Figure \ref{sedfig} shows two example good-fit models whose
parameters are listed in Table \ref{modparms}. 
We found that many 
of the model parameters are quite unconstrained or strongly
dependent on other parameters or on our choice of such basic values as the exact stellar properties and
dust properties.  For example, %our observations are relatively insensitive to the inner and outer disk radii, and
there are the expected degeneracies between stellar luminosity versus extinction, inner disk radius versus minimum grain size,
and between disk mass and grain optical and size properties.  %Resolved imaging is necessary to distinguish between these effects.

\subsection{SSSPM1102}

SSSPM1102-3431 was identified as an M8.5 brown dwarf in the TW Hya association by \citet{scholz05}. 
Photometry and spectroscopy with {\it Spitzer} were reported by \citet{riaz08, morrow08} and 
\citet{luhman10}, and these results are shown in Figure \ref{sedfig}.  The object has a well-determined
parallax distance of 55.2 pc \citep{teix08} and no reported companions.

The two models in Figure \ref{sedfig} illustrate the effect of adding 10\% amorphous
carbon to a nominal silicate grain composition, providing a better fit to the lack of a strong silicate feature in
the {\it Spitzer} IRS data.
Although most of the disk parameters are not well constrained,  we find a fairly
robust lower limit to the  disk mass of a few $\times 10^{-6}$ \msun\ which is driven by the strong far-ir emission  detected.
The probability distribution is quite flat above $10^{-5}$ \msun.
The disk inclination is  also likely greater than 60\degree.

\subsection{ISO138}

ISO138 was identified by \citet{gomez02} as an M5.5 brown dwarf in the Chamaeleon I association, 
though subsequent spectroscopy
by \citet{luhman04} found a spectral type of M6.5.  {\it Spitzer} photometry of ISO138 
reported by \citet{luhman08a} and IRS spectroscopy over the 10\micron\ silicate feature by \citet{pascucci09}
are shown in Figure \ref{sedfig}.  Based on its presumed membership in
Cha I, we assume a distance of 160 pc \citep{luhman08b}, and ISO138 also has no reported companions.

Unlike the other two objects, ISO138 exhibits a weak silicate emission feature \citep{pascucci09}. 
The two model fits shown in Figure \ref{sedfig} illustrate the effect of silicate composition on the shape of the emission
feature.
%ISO138 was the most problematic object to model, possibly due to a combination of its more uncertain
%stellar parameters, e.g. spectral type, and its quite different SED in the 8--70\micron\ spectral region.
The fitted inclination angle and $A_v$ were strongly dependent on the assumed stellar
parameters.  The only fitted parameter for which we have a reasonable constraint is the mass for
which a broadly-peaked probability distribution was found around M$\sim$ few $\times 10^{-6}$ \msun.
The far-ir flux implies a lower disk mass limit of a few $\times 10^{-7}$ \msun.  The inner disk radius is also
constrained to the likely range 0.03--0.15 AU.
%possibly suggesting a larger population of smaller dust grains than for the two TWA BD's.

\subsection{2M1207}

\citet{gizis02} identified 2MASS 1207334-393254 as a likely M8 brown dwarf in TWA
as was subsequently confirmed by \citet{chauvin04}.  A trigonometric parallax distance of 52.4 pc has
been derived by \citet{ducourant08}.  {\it Spitzer} photometry and spectroscopy are described
by \citet{riaz08, morrow08} and are also shown in Figure \ref{sedfig}. 
\citet{riaz08} plot a 70\micron\ MIPS measurement in their Figure 6, but no flux value is described in the
text other than an uncertainty of $\sim$ 0.4 mag, suggesting a 2$\sigma$ detection.
Unlike the other BD's, 2M1207 has a clearly identified companion at a projected separation
of $\sim$ 770 mas (55 AU)\citep{chauvin04,
chauvin05}.  The mass and model-fit to the companion's SED are a subject of some controversy in the
literature, e.g. \citet{mohanty07, mamajek07, skemer11}, but for the purposes of our modeling we 
assume the primary has a substantially higher mass and luminosity than the secondary, and
provides most of the dust heating.

The two models shown in Figure \ref{sedfig} illustrate the negligible effect on the SED of truncating the outer
radius of the disk due to possible effects of the known companion which would be expected to limit the outer
disk radius to $\sim 1/3$ the component separation.
Like SSSPM1102, most disk parameters are quite unconstrained by our models, but the mass and flaring index
do exhibit peaked probability distributions.
Our modeling suggests a likely disk mass $\sim 10^{-5}$ \msun\ with a lower limit of a few $\times 10^{-6}$ \msun,
though masses up to $10^{-4}$ \msun\ 
can also produce reasonable fits.  The flaring index $\gamma$ is fairly well constrained to the range $1.1 < \gamma < 1.15$.
%i.e. a value somewhat less than is typical for more massive objects.  
%The disk scale height parameters are somewhat better
%constrained for this disk than the others above with $h_0 \le 20$ AU at 100 AU, and $1.0 \le \gamma \le 1.15$.
%Most other parameters are rather unconstrained.  
%For example, even though the companion may lead to a truncated
%outer disk, the model SEDs in Figure \ref{sedfig} for two different outer radii show that the observed part
%of the SED is unaffected by the choice of disk outer radius.

\section{Discussion}\label{anal}

%\subsection{Model Results and Comparison to Previous Models}

The strongest conclusion from our modeling is that the addition of the new {\it Herschel} data can provide
important lower limits to the disk masses.   Strong upper limits depend on as yet unavailable photometry at
longer wavelengths and resolved imaging.
%The strongest conclusion from our modeling is that it is very difficult to tightly constrain {\it any} of the disk
%parameters.  
With respect to most of the disk parameters, quite simple assumptions about the disk properties provide good fits to the
observed SED's, but changes to different parameters produce corresponding changes in other fitted 
parameters that still leave a model with a good fit.  Our derived disk masses, for example, suggest
that the SSSPM1102 disk is the most massive, but as for all modeling, this
conclusion depends on assumed grain properties.
%If we make the assumption that all three disks have similar sizes and dust content (composition and
%size distribution), then
%the most certain conclusions about the disks %surrounding these three BD's 
%are the relative masses, i.e.
%the least massive is probably that around ISO138 with those around SSSPM102 and 2M1207 having substantially larger and comparable masses.
With such assumptions, the range from the least to the most massive is likely to lie between a few $\times\ 10^{-6}$ \msun\ up
to perhaps as much as $10^{-4}$ \msun\ (gas$+$dust), though with the most conservative uncertainty estimates, this range could
be 3 times larger or smaller.  The disk scale heights, $h_0$ at 100 AU, likely lie in the range of 5--20 AU
with modest flaring indices, $\gamma < 1.25$, both values that are typical in disk models for more massive objects.
Interestingly the nominally more evolved objects may have less flared disks, based on their differences from ISO138,
but more observations of additional BD's are needed to determine if this is a general effect.
Estimating sub-stellar masses is notoriously uncertain, especially
for very young objects, 
e.g. \citet{baraffe09}. But using the spectral types mentioned above and ages of 1-2 Myr for Cham I and
8 Myr for TWA and common evolutionary BD models, these three objects probably have masses of a 
few $\times$ 0.01 \msun\ \citep{burrows97, baraffe03}.
This would imply a ratio of disk-to-stellar mass in the range  $10^{-4}$ up to a few $\times 10^{-3}$ for these three BD's.
An additional conclusion from our modeling is that it is more difficult to fit the observed SED's with a grain size distribution
appropriate for the interstellar medium than one in which there has been substantial growth in the maximum grain size, 
%with our assumed silicate grain composition, 
but even this conclusion depends somewhat on other assumptions about likely parameter values.
Our modeling shows that the lack of a silicate emission feature in the SED's of 2M1207 and SSSPM1102 can be produced
by a modest admixture of amorphous carbon to the grain composition, though this could also be due to a lack
of small silicate grains.
%This conclusion is driven by the observed 70\micron\ fluxes.  
%Our preliminary models presented here that used pure silicate grains, e.g. the model for 2M1207, do not reproduce
%the lack of a silicate emission feature in the data for 2M1207 and SSSPM102. But 
%%it is clear from our range of
%%attempted models that 
%the simple addition of a small percentage of carbon grains to the mixture can produce a much better fit
%as shown in Figure \ref{sedfig} for SSSPM1102.  Of course, using a different free parameter such as a larger minimum grain
%size could also resolve this issue.

The two of our objects with complete {\it Spitzer} IRS observations have been modeled by \citet{riaz07,riaz08} and \citet{morrow08}.
A number of small differences in model details make a direct comparison difficult.  For example, \citet{riaz08}
used different grain size distributions in the midplane and atmosphere of the disk to simulate dust
settling. \citet{morrow08} included an
expanded ``wall'' at the disk inner edge and characterized their disks with accretion parameters as well as using two 
grain populations.  
Like our models, though, they both assumed a grain size distribution, $n(a) \propto a^{-3.5}$ and were both unable to
constrain the outer disk radius.
\citet{morrow08} do not quote a total disk mass; \citet{riaz08} find a larger mass for the 2M1207 
disk than that for SSSPM1102, but they did not have available the longer wavelength 160\micron\ {\it Herschel} data.
Like our results, both other studies find
evidence for substantial grain growth and for inclinations of $\ge$60\degree.
Most importantly, none of our models required either dust settling nor an expanded inner wall to fit the
observed SED's with reduced $\chi^2 \le$ 1.4.
In general
we find that it is difficult to constrain any of the disk parameters to the degree suggested
by previous modeling without making assumptions that are themselves rather uncertain.
%One difference is that \citet{riaz08} found a substantially higher mass for 2M1207 than SSSPM1102, 
%despite the almost identical fluxes of the two objects over their
%entire SED's.
%but that result appears to have been driven by their low S/N {\it Spitzer} measurement at 70\micron.  
%They also found a fairly high
%inclination angle for 2M1207 of 75\degree\ based on detection of silicate absorption in their IRS data, but
%\citet{morrow08} did not claim to see any silicate feature, either in emission or absorption, in the same data set but
%using a different reduction.

%As is true for most modeling of thermal dust emission, 
%The most efficient way to further constrain
%SED-derived models is to obtain spatially resolved imagery at several wavelengths, e.g. near-infrared and
%sub-mm interferometry.  
To further constrain these models the next obvious step is to extend the SED's to $\lambda \sim$ 1 mm and obtain spatially
resolved images with ALMA.
The brighter members of our BD sample are likely to be resolvable with the full
ALMA array.
The available model parameter space may also be better constrained with the highest S/N
spectroscopy to best define the sub-stellar photospheres, in particular for ISO138 which has discrepant spectral
types in the literature.
%The final determination of accurate disk parameters, however,  
%depends to a great deal on the kind of model used,
%the assumptions about underlying BD photospheric parameters, and the signal-to-noise in the observations.

One other BD has been observed over an even greater span of the electromagnetic spectrum, 2M04442713+2512
\citep{bouy08} from the visible to 3.5mm.   They inferred a 
fairly massive disk, $\sim 10^{-3}$ \msun, though
even with their complete and high S/N observations, many disk parameters were still poorly constrained.
Disk masses for BD's have also been derived for objects with millimeter wavelength photometry by \citet{klein03}
and \citet{scholz06}.  They estimated masses in the range of $\sim 10^{-6}$--$10^{-3}$ \msun\ with differing
assumptions, but clearly in the same range as studies based on more complete SED's.
It is also interesting to compare these four BD's, i.e. our three plus 2M04442713, to T Tauri stars.
% which are
%roughly the more massive analogs to our young BD's and for which a great deal more and higher
%quality data exist.  For example, 
\citet{andrews05, andrews07} have surveyed T Tau stars at millimeter
wavelengths where the dust is optically thin and the most accurate masses can be determined.
They find {\it dust} masses in the range of $10^{-5}$ up to perhaps as high as $10^{-3}$ \msun, depending on
assumed grain sizes and properties, implying total disk masses of $10^{-3}$ -- $10^{-1}$ \msun\ with our
canonical gas-to-dust ratio.   So the low end of the T Tau disk mass distribution overlaps with the
high end, 2M04442713, of the BD distribution with the very small sample available so far for BD's.
Recently \citet{lee11} have found substantially lower disk masses around T Tauri stars in the 2 Myr old
IC 348 association than in the very youngest, nearby star-forming clouds, Taurus and Ophiuchus.  Since
TWA and probably Cham I are both older than Taurus and Ophiuchus, it may be best to compare our BD disks
to T Tauri disks in older star-forming clouds.  With the completion of our data set within the coming
year, we should have enough statistics to discern any such trends with better significance.
%The low end of our limited BD disk mass distribution is comparable to the most massive debris disks around
%more massive main-sequence stars, but the BD disk SED's are clearly more like those of T Tauri stars.

\section{Summary}\label{summ}

We have detected all three of our first program objects at 70\micron\ and one at 160\micron.
These observations represent by far the most
sensitive far-infrared photometry of brown dwarfs.
Our modeling shows that the SED's can be fit with simple geometric disks that do not require
an inner wall nor dust settling.
The addition of our Herschel measurements provides much stronger lower limits to the masses of
the circumstellar disks because most of the disk mass is at large radii and relatively cool.
The implied disk masses are probably well below those surrounding many of their more
massive counterparts, the T Tau stars.
These disks are likely to be
optically thin in the far-ir perpendicular to the disk plane outside 1 AU, though optical depths through the midplane are still 
high, $A_v \ge 1000$.
Interestingly the least massive disk is found around the nominally youngest BD, ISO138 in Cham I, while
the disks around the BD's in TWA, presumed to have an age of order 8 Myr, are probably more massive.
Future ALMA observations will enable more accurate masses and disk sizes to be determined for many of
the objects in our sample that are bright enough to be resolved.  For example, the most massive disk 
in this small sub-sample, that around
SSSPM1102, would likely have a flux of slighly over 1 mJy at 850\micron, and these three objects are
some of the fainter members of our total sample.

\section{Acknowledgments}
Support for this work, as part of the NASA Herschel Science Center data analysis funding program, 
was provided by NASA through a contract issued by the Jet Propulsion Laboratory, California 
Institute of Technology to the University
of Texas.  LAC was supported by NASA through the Sagan Fellowship Program.
FM  and CP acknowledge support from ANR (contract ANR-07-BLAN-0221 and ANR-2010-JCJC-0504-01), European Commission's
7$^{\mathrm(th)}$ Framework Program (contract PERG06-GA-2009-256513), and Programme National de Physique 
Stellaire (PNPS) of CNRS/INSU, France.
SW acknowledges support by the German Research Foundation
(contract FOR 759). YL acknowledges support by the
German Academic Exchange Service.

\clearpage

\begin{table}[h]
\caption{Observational Parameters \label{obstable}}
\vspace {3mm}
\begin{tabular}{lccc}
\tableline
\tableline
Parameter  & Value  &  Comments \cr
\tableline

%\bigskip

AOR Type &  PACS Mini-Scan-Map &  Two Crossed AOR's \\
Wavelengths &  70\micron,  160\micron\ & \\
Number of Scan Legs &  8 & \\
Scan Length &  3\arcmin\ & \\
Cross Scan Step & 4\arcsec\ & \\
Scan Angles &  70\degree, 110\degree & Relative to Detector \\
Repetitions &   7  & Per AOR \\
Peak Intg Time Per Pixel & 504 sec & Per AOR \\

\tableline

\tableline
\end{tabular}
\end{table}

%\begin{table}[h]
%\caption{Observations Summary (Program ID = GT1\_pharve01\_2) \label{aortable}}
%\vspace {3mm}
%\begin{tabular}{lcccc}
%\tableline
%\tableline
%Object & RA/Dec Center (2000) & AOR's  & Obs. Date  \cr
%\tableline
%
%       SSSPM1102 & 11 02 09.8 -34 30 36 & 1342221849/50 & 29 May 2011 \cr
%       ISO138 & 11 08 19.0 -77 30 41 & 1342218699/700 & 16 Apr 2011 \cr
%2MASS1207334-393254 & 12 07 33.4 -39 32 54 & 1342202557/58 & 10 Aug 2010 \cr
%\tableline

%%\bigskip

%\tableline

%\tableline
%\end{tabular}
%\end{table}

%\begin{table}[h]
%\caption{Observed Flux Densities  \label{fluxtable}}
%\vspace {3mm}
%\begin{tabular}{lcc}
%\tableline
%\tableline
%Object & 70\micron\ (mJy)  & 160\micron\ (mJy)  \cr
%\tableline

%       SSSPM1102 & 7.3 $\pm$ 1.0 & 7.1 $\pm$ 1.5 \cr
%       ISO138 & 3.7 $\pm$ 0.6 & $<$ 15 \cr
%2MASS1207334-393254 & 7.0 $\pm$ 0.8 & $<$ 7 \cr
%\tableline

%%\bigskip

%\tableline

%\tableline
%\end{tabular}
%\end{table}

\begin{table}[h]
\caption{Observations Summary (Program ID = GT1\_pharve01\_2) \label{aortable}}
\vspace {3mm}
\begin{tabular}{lccccc}
\tableline
\tableline
Object & RA/Dec Center (2000) & AOR's  & Obs. Date & 70\micron\ (mJy)  & 160\micron\ (mJy)  \cr
\tableline

       SSSPM1102 & 11 02 09.8 -34 30 36 & 1342221849/50 & 29 May 2011 & 7.3 $\pm$ 1.0 & 7.1 $\pm$ 1.5 \cr
       ISO138 & 11 08 19.0 -77 30 41 & 1342218699/700 & 16 Apr 2011& 3.7 $\pm$ 0.6 & $<$ 15  \cr
2M1207 & 12 07 33.4 -39 32 54 & 1342202557/58 & 10 Aug 2010 & 7.0 $\pm$ 0.8 & $<$ 7 \cr
\tableline

%       SSSPM1102 & 7.3 $\pm$ 1.0 & 7.1 $\pm$ 1.5 \cr
%       ISO138 & 3.7 $\pm$ 0.6 & $<$ 15 \cr
%2MASS1207334-393254 & 7.0 $\pm$ 0.8 & $<$ 7 \cr
\tableline

\end{tabular}
\end{table}

%\clearpage
\begin{deluxetable}{lccc}%[h]
%\rotate
\tablecolumns{4}
\tablecaption{Parameters (and Range Explored) For Models in Figure \ref{sedfig}\label{modparms}}
\tablewidth{0pt}
%\vspace {3mm}
%\begin{tabular}{lccc}
%\tableline
%\tableline
\tablehead{
%Model & Parameter  & Value  &  Comments \cr
\colhead{}  &  
\colhead{ISO138} &
\colhead {2M1207} & 
\colhead{SSPM1102} \\
%\tableline
\colhead{Parameter} &
\colhead {\tablenotemark{1}Value (Range)} &
\colhead{\tablenotemark{1}Value (Range)} & 
\colhead {\tablenotemark{1}Value (Range)} 
}
%\tableline
\startdata

{\bf Stellar} \cr
$T_{eff}$ (K) & 2900 & 2600 &  2600 \cr 
$R_{star}$ (\rsun) & 0.35 (0.2--0.5) & 0.24 (0.2--0.3) &  \tablenotemark{a}0.35, \tablenotemark{b}0.27 (0.2--0.4) \cr 
Luminosity (\lsun) & 0.0168 & 0.0046 & 0.0059 \cr
%\smallskip
$A_v$ &  0.5 & 0.0 & \tablenotemark{a}0.0, \tablenotemark{b}1.5 \cr
Dist. (pc) & 160 & 53 & 56 \cr
%\smallskip
{\bf Disk} \cr
Incl. (deg) & $<$60 (0--90) &  \tablenotemark{a}70, \tablenotemark{b}78 (0--90) &  \tablenotemark{a}80, \tablenotemark{b}66 (0--90) \cr
$R_{inner}$ (AU) & 0.08 (0.08--40) & 0.015 (.005--.03) & \tablenotemark{a}0.015, \tablenotemark{b}0.006 (.0035--.015) \cr
$R_{outer}$ (AU) & 100 & \tablenotemark{a}75, \tablenotemark{b}20 &  75 \cr
$H_o$(AU)@100AU & 20 (5--30) & 9 (5--15) & 5 (5--15)  \cr
$\gamma$ & 1.25 (1.0--1.25) &  1.125 (1.00--1.125) & \tablenotemark{a}1.05, \tablenotemark{b}1.07 (1.0-1.125) \cr
-p & 0.8 (0.5--1.5) & 1.0 (0.5--1.5) & \tablenotemark{a}0.5, \tablenotemark{b}0.1 (0.1--1.5) \cr
%\smallskip
{\bf Dust} \cr
%$M_{dust}$ (\msun) & 2$\times10^{-8}$,5.0$\times10^{-8}$ \cr
%$M_{dust}$ (\msun) & 2.0e-08, 5.0e-08 (2e-09--2e-06) & 2e-07, 5e-06 (????) & ??, 5e-07 (???) \cr
\tablenotemark{2}Log($M_{disk})$ (\msun) & -5.2 (-6.7-- -3.7) & \tablenotemark{a}-5.0 \tablenotemark{b}-5.2 (-6.5-- -3.5) & -4.0 (-6.5-- -3.5) \cr
%\smallskip
$a_{min}$(\micron) & 0.05 &  0.1 & 0.05 \cr
$a_{max}$(\micron) &  1000 (10--1000) & 1000 (10--1000) &  10 (10--1000) \cr
Power law & -3.5 & -3.5 & -3.5 \cr
Silicate & \tablenotemark{a}0.95 DL, \tablenotemark{b}0.95 Olivine &  0.95 DL  &  \tablenotemark{a}0.90 DL, \tablenotemark{b}1.00 DL \cr
Amorphous Carbon & 0.05  &   0.05 & \tablenotemark{a}0.10, \tablenotemark{b}0.00 \cr

% \\

%\tableline

%\tableline
%\end{tabular}
\enddata
\tablenotetext{1}{Where two values are given, the first is for the ``a'' model and the second for the ``b'' model. Otherwise 
parameters were the same for both models.}
\tablenotetext{2}{Assuming gas-to-dust mass ratio of 100.}
\end{deluxetable}

%%%%%%%%%%%%%%%%%% Tables %%%%%%%%%%%%%%%%%%%%%

\clearpage
%%%%%%%%%%%%%%%%%% Figures %%%%%%%%%%%%%%%%%%%%%

\clearpage

\end{document}